# A novel method based solely on FPGA units enabling measurement of time and charge of analog signals in Positron Emission Tomography

M. Pałka[1], T. Bednarski[1], P. Białas[1], E. Czerwiński[1], Ł. Kapłon[1,2], A. Kochanowski[2], G. Korcyl[1], J. Kowal[1], P. Kowalski[3], T. Kozik[1], W. Krzemień[1], M. Molenda[2], P. Moskal[1], Sz. Niedźwiecki[1], M. Pawlik[1], L. Raczyński[3], Z. Rudy[1], P. Salabura[1], N.G. Sharma[1], M. Silarski[1], A. Słomski[1], J. Smyrski[1], A. Strzelecki[1], W. Wiślicki[3], M. Zieliński[1], N. Zoń[1]

[1]Institute of Physics, Jagiellonian University, 30-059 Cracow, Poland
[2]Faculty of Chemistry, Jagiellonian University, 30-060 Cracow, Poland
[3]Świerk Computing Centre, National Centre for Nuclear Research, Otwock-Świerk, Poland

**Abstract:**
This article presents a novel technique for precise measurement of time and charge based solely on FPGA (**F**ield **P**rogrammable **G**ate **A**rray) device and few satellite discrete electronic components used in **P**ositron **E**mission **T**omography (PET). Described approach simplifies electronic circuits, reduces the power consumption, lowers costs, merges front-end electronics with digital electronics and also makes more compact final design. Furthermore, it allows to measure time when analog signals cross a reference voltage at different threshold levels with a very high precision of ~10ps (rms) and thus enables sampling of signals in a voltage domain.

**Introduction:**
This article is focused on novel measurement methods of analog signals and its electrical parameters in TOF-PET tomography [1,2,3]. Positron Emission Tomography is used to determine the spatial and temporal density distribution of selected substances in the body. PET detectors register gamma quanta originating from the annihilation of positrons with electrons. Positrons are emitted by the radio-isotopes included in the radiopharmaceuticals administered to the patient before the diagnosis.

The core of the PET scanner comprises scintillators converting the energy of annihilation quanta into light pulses which are subsequently converted to electrical signals by means of photomultipliers or photo-diodes. To measure time and charge of such signals one has to use set of electronic circuits, which allow to discriminate signals (analog discriminators) and convert time to digital information (**T**ime to **D**igital **C**onverter - TDC). Currently known discriminators are commonly used in area of experimental physics or in devices for medical diagnostics. Analog discriminators can be divided into: constant-level and constant-fraction. In connection with TDC they allow measurement of the time at which the signal exceeds the preset voltage or a desired fraction of the amplitude, respectively. They play a particularly important role in the latest generation of positron emission tomographs (so called TOF-PET) which utilize information about a difference in Time-Of-Flight (TOF) between the annihilation point and the detector for the two registered gamma quanta.

Typical discriminators are built on the basis of standard electronic components and include among other things: current source, preamplifier, comparator, shaper, capacitors, resistors, diodes, transistors and transmission lines. In case of TDC a dedicated ASICs (**A**pplication **S**pecific **I**ntegrated **C**ircuit) is used.

In this article an utterly novel solution is presented which allows to incorporate discrimination and TDC into FPGA unit. For this purpose **L**ow **V**oltage **D**ifferential **S**ignal (LVDS) buffers of the FPGA device are "misuse" as a comparator and internal carry-chain elements as delay elements.

**FPGA-TDC:**

Time measurement technique in FPGAs is based on the possibility of using carry-chains usually used as part of adders [4,5]. Carry-chain lines, because of its nature (adding long words), have minimized delay (in order to increase the possible maximum operating frequency of the FPGA).

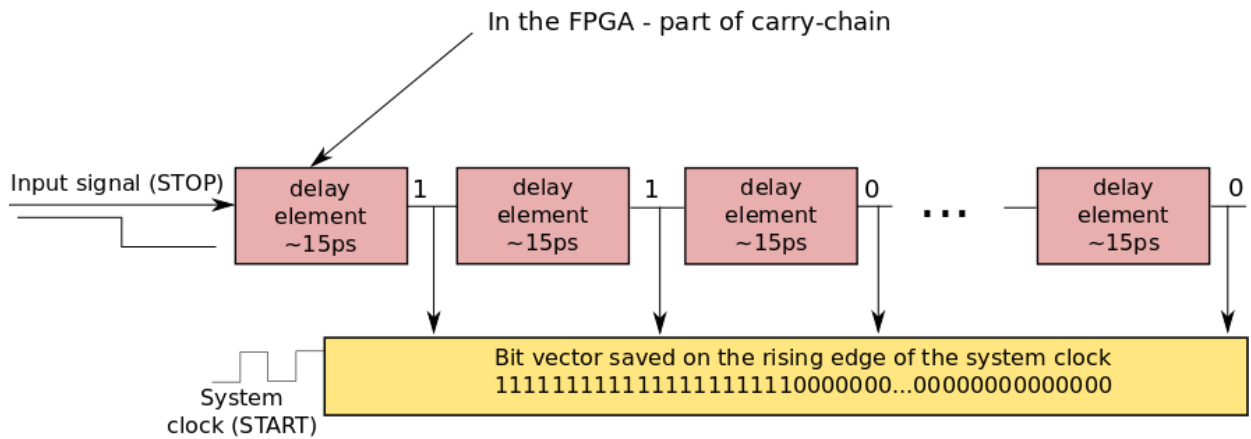

**Figure 1**: Diagram of time measurement when using a carry-chain as a delay line.

As shown in Fig. 1 the time measurement is based on writing a bit vector, which represents STOP signal (measured signal), in the D-type flip-flops with the rising edge of the START signal (system clock). STOP signal is delayed by the elements of carry chain, which normally is used by adders in a FPGA. The lower latency of a single element of the carry-chain translates to better precision of a time measurement.

In order to achieve high precision of time measurement one should take into account corrections related with the non-linearity of measurement (differential and integral nonlinearity) as well as the impact of temperature and supply voltage (measurable directly in the FPGA). These corrections can be added off-line or in real time to the measured values. Another difficulty is the so-called meta-stability resulting from different reaction time to the transition of the signal. When saving the current state of the STOP signal it may be a case when there is no continuity of ones and zeros on the rising/falling edge of the measured signal (Fig. 2). Discontinuity of ones and zeros can also result from the different path lengths of carry-chain to the flip-flops. This, however, can be controlled by writing the appropriate limitations relating to the location of the individual elements in the FPGA. As a result, in order to correctly decode binary sequence to time, the required amount of logic must be increased.

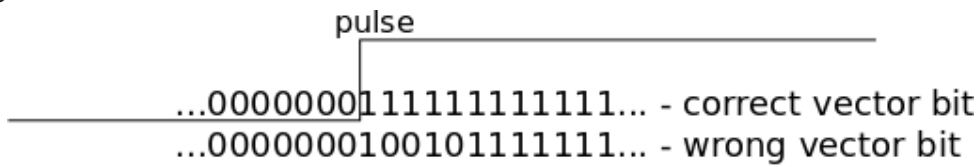

**Figure 2**: Saved value in flip-flops may be not correct due to meta-stability. Therefore the amount of logic required to decode the time increases.

First mentioned TDC technique was already implemented in a Trigger Readout Board version 3 [7]. As it is shown in Fig. 3 it consists of five Lattice ECP3-150 FPGAs and there are no dedicated Time to Digital Converter (TDC) components (like HPTDC ref.[8]). Therefore four of them are foreseen to perform time to digital conversion (TDC) on incoming signals. Board serves 256 TDC channels with ~10ps time resolution. The fifth FPGA is responsible for management of

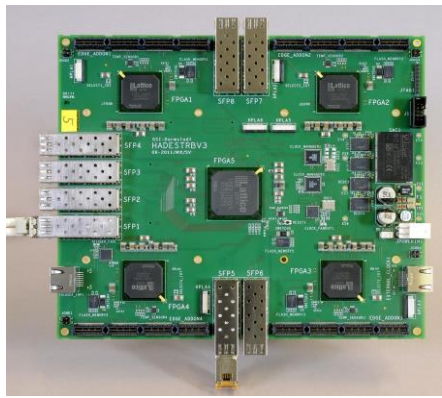

**Figure 3** : TRBv3 for time measurements. Four Lattice FPGAs placed in the corner of the board are used for time measurements. The central one is used to data flow management

**FPGA LVDS input buffer as a discriminator:**

FPGA LVDS buffer is typically used as buffer for digital signal transmission. When positive signal is higher than negative it gives logical '1' and in opposite case '0'. This kind of buffer has to cope with high bit rates (Gbit/s). It should be noted that the input voltage range is limited to the acceptable voltage of FPGA LVDS buffer (usually from 0 to ~2V). These working parameters allow to use it as a discriminator for TOF-PET signals.

In order to verify the applicability of FPGAs as discriminator a preliminary measurements were made. In Fig. 5 a block diagram of the measurement setup is shown. On one of the LVDS buffer differential inputs (+) has been given the reference voltage (25 ns rising edge) and to the second constant level voltage was applied. On the second LVDS buffer a reference signal was connected. It was indicating the beginning of the reference signal. In the FPGA two TDC channels were implemented. With the increase of the voltage level the time difference between the reference signal and the measured signal should also increase (Fig.5). Measured distribution of the time difference corresponds to the achievable discriminator resolution.

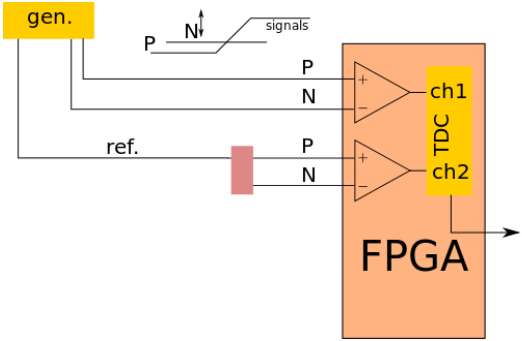

**Figure 4** : Test setup for measuring ADC performance.

| N level [mV] | Time [ps] | Jitter (RMS) [ps] |
|---|---|---|
| 400 | 2551 | 50 |
| 800 | 3034 | 24.00 |
| 1200 | 3538 | 18.00 |
| 1600 | 4125 | 15.00 |

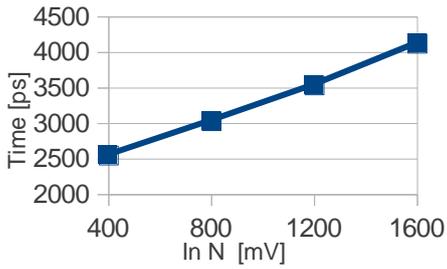

**Table** 1 : Results from first ADC measurements

**Figure 5** : Measured time and input voltage

relationship.

It should be emphasized that the results were achieved without any calibration, whether of non-linearity or to compensate the switching time of the LVDS buffer. The time measurement precision achievable with such LVDS buffers together with FPGA-TDC is sufficient for needs of the TOF-PET tomography.

**FPGA-TDC and FPGA-discriminator in TOF-PET application**

When combining FPGA-TDC and FPGA-discrimination one can sample incoming signal with very high precision (10ps rms) which is particularly important in case of TOF-PET based on the plastic scintillators [1,2,3] with the time of the rising edge of registered signals is in the range of few ns. In order to perform sampling the input signal is split into several ones which are treated independently. The thresholds for the split signals can be adjusted depending on different conditions (base line level and maximum signal amplitude, see Fig. 7). Sampling may improve the time resolution significantly. This may be achieved for example by fitting a function describing the shape of the signals to sampled points (see Fig.7 red line) with the beginning of the measured signal as a free parameter of the fit. In addition, combination of rising and falling edges should allow for measurement of charge with very good resolution. Charge measurement can be later on used for rejection of noise originating from registration of gamma quanta scattered in the body of the patient.

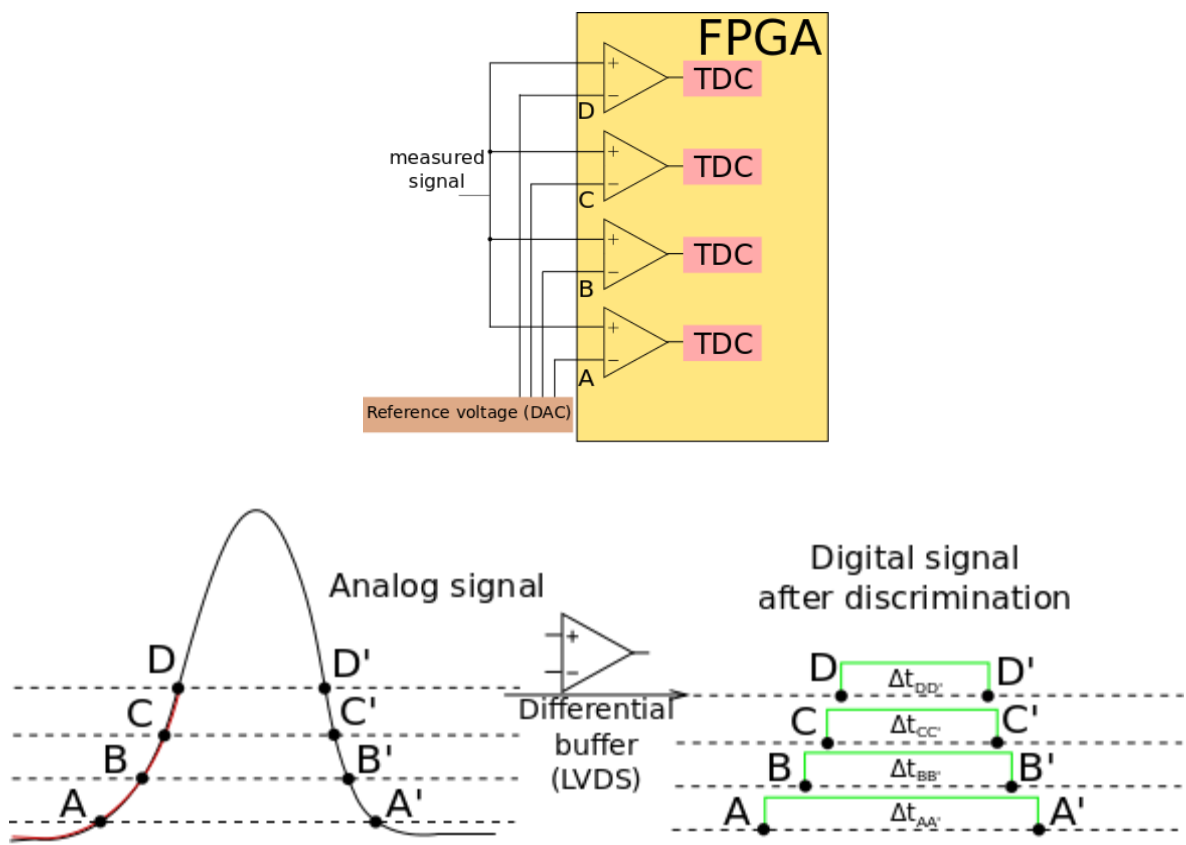

**Figure 6** : An scheme of a four threshold device for the sampling in the voltage domain and the pictorial explanation of its functioning. ADC and TDC mixture increases resolution of time and amplitude measurement. The number of threshold can be increased according to the needs.

**Conclusions and Discussion**

Currently TDC and QDC (charge-**Q** to **D**igital **C**onverter) devices are widely used in the research as well as in the various types of diagnostic devices. Typically time measurement is performed by dedicated ASICs [7] with the precision of up to about 20ps (rms) and in case of charge an accuracy of less than 1% is achievable. However, in these cases, all time circuits (converters) have to be read-out and their parameters have to be controlled by the peripheral devices. Most often this is done by using microcontrollers or FPGA's.

In this article a novel method of the time measurement was presented which allows for sampling of analog signals in the voltage domain with a precision of 10ps (rms). The method is based on the usage of LVDS buffers of the FPGA device as a comparator and internal carry-chain elements as delay units. It was developed for the purpose of sampling of fast signals in the newly developed TOF-PET detector system based on the plastic scintillators.

The benefits from presented measurement techniques integrated within FPGA are:
- Reduction of cost of measurement systems - no need to purchase a dedicated ASIC,
- Low power consumption (only FPGA and discrete components),
- Much higher concentration of measurement channels compared to the traditional approach,
- Much more flexibility and possibility of miniaturization of the measurement systems.

**Acknowledgements**

We acknowledge technical and administrative support by M. Adamczyk, T. Gucwa-Ryś, A. Heczko, M. Kajetanowicz, G. Konopka-Cupiał, J. Majewski, W. Migdał, A. Misiak and the financial support by the Polish National Center for Development and Research through grant INNOTECH-K1/IN1/64/159174/NCBR/12, the Foundation for Polish Science through MPD programme and the EU and MSHE Grant No. POIG.02.03.00-161 00-013/09.